\documentclass[12pt]{article}

\usepackage{epsf}
\usepackage{graphicx}
\usepackage[figuresright]{rotating}

\newcommand\lsim{\mathrel{\rlap{\lower4pt\hbox{\hskip1pt$\sim$}}
    \raise1pt\hbox{$<$}}}
\newcommand\gsim{\mathrel{\rlap{\lower4pt\hbox{\hskip1pt$\sim$}}
    \raise1pt\hbox{$>$}}}
\def\lsim{\mathrel{\raise.3ex\hbox{$<$\kern-.75em\lower1ex\hbox{$\sim$}}}} 
\def\gsim{\mathrel{\raise.3ex\hbox{$>$\kern-.75em\lower1ex\hbox{$\sim$}}}}

\def\beq{\begin{equation}}
\def\eeq{\end{equation}}
\def\ba{\begin{eqnarray}}
\def\ea{\end{eqnarray}}
\newcommand{\tb}{\tan \beta}
\def\li#1{\iso{Li}{#1}}
\def\he#1{\iso{He}{#1}}
\newcommand\iso[2]{\mbox{${}^{#2}${\rm #1}}}

\begin{document}
\begin{titlepage}
\pagestyle{empty}
\baselineskip=21pt
\begin{center}
{\Large {Supersymmetric Dark Matter Candidates}}
\end{center}
\begin{center}
{{John~Ellis}~$^a$ and
{Keith~A.~Olive}~$^b$}\\
\vskip 0.05in
{\it \small
$^a${TH Division, Physics Department, CERN, Geneva, Switzerland}\\
$^b${William I.\ Fine Theoretical Physics Institute,
University of Minnesota, Minneapolis, MN~55455, USA}\\
}
\vskip 0.1in
{\bf Abstract}
\end{center}
\baselineskip=18pt \noindent
\vskip -2.5in
{\rightline{\small CERN--PH--TH/2010-004, UMN--TH--2831/10, FTPI--MINN--10/02}}
\vskip 2.2in
{\small
After reviewing the theoretical, phenomenological and
experimental motivations for supersymmetric extensions of the
Standard Model, we recall that supersymmetric relics from the Big Bang are
expected in models that conserve $R$ parity. We then discuss possible
supersymmetric dark matter candidates, focusing on the lightest neutralino 
and the gravitino. In the latter case, the next-to-lightest supersymmetric
particle is expected to be long-lived, and possible candidates include
spartners of the tau lepton, top quark and neutrino. We then discuss the
roles of the renormalization-group equations and electroweak symmetry
breaking in delimiting the supersymmetric parameter space. We discuss
in particular the constrained minimal extension of the Standard Model
(CMSSM), in which the supersymmetry-breaking parameters are assumed to
be universal at the grand unification scale, presenting predictions from a
frequentist analysis of its parameter space. We also discuss astrophysical
and cosmological constraints on gravitino dark matter models, as well as
the parameter space of minimal supergravity (mSUGRA) models in which
there are extra relations between the trilinear and bilinear supersymmetry-breaking
parameters, and between the gravitino and scalar masses. Finally, we discuss
models with non-universal supersymmetry-breaking contributions to Higgs masses,
and models in which the supersymmetry-breaking parameters are
universal at some scale below that of grand unification.}

\begin{center}
{\it \small From  `Particle Dark Matter: Observations, Models and Searches'
edited by Gianfranco Bertone
Copyright © 2010 Cambridge University Press.
Chapter 8, pp. 142-163
Hardback  ISBN  9780521763684, 
http://cambridge.org/us/catalogue/catalogue.asp?isbn=9780521763684}
\end{center}



\end{titlepage}
\baselineskip=18pt






\section{Motivations}

Supersymmetry is one of the best-motivated proposals for physics
beyond the Standard Model. There are many idealistic motivations for 
believing in supersymmetry, such as
its intrinsic elegance, its ability to link matter particles and force carriers, its
ability to link gravity to the other fundamental interactions, its essential role
in string theory, etc. However, none of these aesthetic motivations gives any
hint as to the energy scale at which supersymmetry might appear.
The following are the principal utilitarian reasons to think that supersymmetry
might appear at some energy accessible to forthcoming experiments. 

The first and primary of
these was the observation that supersymmetry could help stabilize the mass
scale of electroweak symmetry breaking, by cancelling the quadratic
divergences in the radiative corrections to the mass-squared of the Higgs boson
\cite{Maiani:1979cx,'tHooft:1980xb,Witten:1981kv},
and by extension to the masses of other Standard Model particles. This
motivation suggests that sparticles weigh less than about 1~TeV, but the exact 
mass scale depends on the amount of fine-tuning that one is prepared to
tolerate. 

Historically, the second motivation for low-scale supersymmetry,
and the one that interests us most here,
was the observation that the lightest supersymmetric particle (LSP) in
models with conserved $R$ parity, being heavy and
naturally neutral and stable, would be an excellent candidate for dark
matter~\cite{Goldberg:1983nd,Ellis:1983ew}. This motivation requires that the lightest
supersymmetric particle should weigh less than about 1~TeV, if it had once
been in thermal equilibrium in the early Universe. This would have been the
case for a neutralino $\chi$ or a sneutrino $\tilde \nu$ LSP, and the argument can
be extended to a gravitino LSP because it may be produced in the decays of
heavier, equilibrated sparticles.

The third reason that emerged for thinking that supersymmetry may be accessible 
to experiment was the observation that including sparticles in the 
renormalization-group equations (RGEs) for the gauge couplings of the Standard Model
would permit them to unify
\cite{Ellis:1990zq,Ellis:1990wk,Amaldi:1991cn,Langacker:1991an,Giunti:1991ta}, 
whereas unification would not occur if only the
Standard Model particles were included in the RGEs. However, this argument does not
constrain the supersymmetric mass scale very precisely: scales up to about
10~TeV or perhaps more could be compatible with grand unification.

The fourth motivation is the fact that the Higgs boson is (presumably)
relatively light, according to the precision electroweak data - an argument
reinforced by the negative results (so far) of searches for the Higgs boson
at the Fermilab Tevatron collider. It has been known
for some 20 years that the lightest supersymmetric Higgs boson should weigh
no more than about 140~GeV, at least in simple 
models~\cite{Ellis:1990nz,Ellis:1991zd,Okada:1990gg,Okada:1990vk,Haber:1990aw}.
Since the early 1990s,
the precision electroweak noose has been tightening, and the best indication now
(incorporating the negative results of searches at LEP and the Tevatron) is that
the Higgs boson probably weighs less than about 140~GeV~\cite{EWWG,TevEWWG}, 
in perfect agreement with the supersymmetric prediction.

Fifthly, if the Higgs boson is indeed so light, the present electroweak vacuum would be
destabilized by radiative corrections due to the top quark, unless the Standard Model is 
supplemented by additional scalar particles~\cite{Ellis:2000ig}. This would be automatic in supersymmetry,
and one can extend the argument to `prove' that any mechanism to stabilize the electroweak 
vacuum must look very much like supersymmetry.

There is a sixth argument that is still controversial, namely the anomalous magnetic moment of the
muon, $g_\mu - 2$. As is well known, the experimental measurement of this quantity~\cite{Bennett:2006fi} disagrees with the Standard Model prediction~\cite{Passera:2008jk}, 
if this is calculated using low-energy $e^+ e^-$
annihilation data. On the other hand, the discrepancy with the Standard Model is greatly
reduced if one uses $\tau$ decay data to estimate the Standard Model contribution to
$g_\mu - 2$. Normally, one would prefer to use $e^+ e^-$ data, since they are related more
directly to $g_\mu - 2$, with no need to worry about isospin violations, etc. 
Measurements by the BABAR collaboration using the radiative-return 
method~\cite{Davier:2009zi} 
yield a result intermediate between the previous $e^+ e^-$ data and $\tau$ decay data. 
Until the discrepancy between these data sets have been ironed out, one should take 
$g_\mu - 2$ {\it cum grano salis}.

\section{The MSSM and $R$ parity}

We refer to~\cite{Fayet:1976cr,Nilles:1983ge} for the general structure of supersymmetric
theories. We restrict ourselves here to theories with a single supersymmetry
charge, called simple or $N=1$ supersymmetry, as these are the only ones able
to accommodate chiral fermions and hence the violation of parity and charge conjugation.
We recall that the basic building building
blocks of $N=1$ supersymmetric models are so-called chiral supermultiplets, each
consisting of a Weyl fermion and a complex scalar, and gauge supermultiplets,
each consisting of a gauge field and a gaugino fermion. The renormalizable 
interactions between the chiral supermultiplets are characterized by a superpotential
that couples the chiral supermultiplets in bilinear and trilinear combinations
that yield masses and Yukawa interactions, and
by gauge interactions. In this framework,
bosons and fermions must appear in pairs with identical internal quantum numbers.
Since the known particles do not pair up in this way, it is necessary to postulate
unseen particles to partner those known in the Standard Model. 

In order to construct the minimal supersymmetric extension of the Standard
Model (MSSM)~\cite{Fayet:1976et,Fayet:1977yc,Fayet:1979sa}, one starts with the
complete set of chiral fermions needed in the Standard Model, and adds a complex scalar
superpartner to each Weyl fermion, so that each matter field in the Standard Model
is extended to a chiral supermultiplet. These are denoted by $L^i, Q^i, e^c, d^c$ and $u^c$,
where $i, j$ are SU(2)$_L$ doublet indices and generation indices
have been suppressed as were color indices for the quarks. In order to avoid a triangle anomaly, Higgs
supermultiplets must appear in pairs with opposite hypercharges, and the minimal possibility
is a single pair $H_1^i, H_2^i$. One must also add a
gaugino for each of the gauge bosons in the Standard Model so as to complete the
gauge supermultiplets. The minimal supersymmetric standard model (MSSM)~\cite{Haber:1984rc}
is defined by this minimal field content and the minimal superpotential necessary to account for
the necessary Yukawa couplings and mass terms, namely:
\beq
W = \epsilon_{ij} \bigl( y_e H_1^j  L^i e^c + y_d H_1^j Q^i d^c + y_u H_2^i
Q^j u^c \bigr) + \epsilon_{ij} \mu H_1^i H_2^j .
\label{WMSSM}
\eeq 
In (\ref{WMSSM}), the Yukawa couplings, $y$, are all $3 \times 3$ matrices in generation space,
with no generation indices for the Higgs multiplets. A second reason for requiring
two Higgs doublets in the MSSM is that the superpotential must be a holomorphic function of the
chiral superfields.  This implies that there would be no way to account for all of the
Yukawa terms for both up- and down-type quarks, as well as charged leptons, with a single Higgs
doublet.  The physical Higgs
spectrum then contains five states: two charged Higgs bosons $H^\pm$, two scalar neutral
Higgs bosons $h, H$, and a pseudoscalar Higgs boson $A$.
The final bilinear mixing term in (\ref{WMSSM}) must be
included in the superpotential, in order to avoid a massless Higgs state.

The MSSM must be coupled to gravity, which requires the introduction of a graviton
supermultiplet containing a spin-3/2 gravitino as well as the spin-2 graviton itself,
which may or not be coupled minimally to the MSSM. The consistency of
supergravity at the quantum level requires the breaking of supersymmetry to be
spontaneous, with the gravitino mass acting as an order parameter \cite{Cremmer:1978hn,
Cremmer:1978iv}. The mechanism
whereby supersymmetry is broken is unknown, as is how this feeds into the MSSM.
We adopt here a phenomenological approach, parametrizing the results of this
mechanism in terms of differing amounts of explicit
supersymmetry breaking in the masses and couplings of the unseen
supersymmetric partners of Standard Model particles 
\cite{Barbieri:1982eh,Arnowitt:1983ah,Nilles:1983ge}. 

In order to preserve the hierarchy between the
electroweak and GUT or Planck scales, it is necessary that this explicit
breaking of supersymmetry be `soft', i.e., in such a way that the theory remains
free of quadratic divergences, which is possible with the insertion of 
weak scale mass terms in the Lagrangian~\cite{Girardello:1981wz}. 
The possible forms for such terms are
\ba
{\cal L}_{soft} & = & -{1\over 2} M^a \lambda^a \lambda^a
-{1\over 2} ({m^2})^{\alpha}_{\beta} \phi_\alpha {\phi^\beta}^* \nonumber \\
& & -{1\over 2} {(BM)}^{\alpha\beta} \phi_\alpha \phi_\beta - {1\over 6} 
{(Ay)}^{\alpha\beta\gamma} \phi_\alpha
\phi_\beta \phi_\gamma + h.c.
\ea 
\noindent where the $M^a$ are masses for the gauginos $\lambda^a$, 
$m^2$ is a matrix of soft scalar masses-squared that carries two field indices, $\alpha, \beta$,
for scalars $\phi_\alpha$,  
$A$ is a trilinear coupling term with three field indices, and $B$ is a bilinear
supersymmetry breaking term associated with a superpotential bilinear mass term
such as $\mu$ in Eq. \ref{WMSSM}.
Masses for the gauge bosons are, as usual, induced by the spontaneous breaking of
gauge invariance, and the masses for chiral fermions are induced by the Yukawa
superpotential terms when the electroweak gauge symmetry is broken. 
For a more complete discussion of supersymmetry and the construction of the MSSM 
see~\cite{Martin:1997ns,Ellis:1998eh,Olive:1999ks,Peskin:2008nw}.

In defining the MSSM, we have limited the model to contain a minimal field
content: the only new fields are those which are {\em required} by
supersymmetry. Consequently, apart from superpartners, only the
Higgs sector was enlarged from one doublet to two. Moreover, in writing the
superpotential (\ref{WMSSM}), we have also made a minimal choice regarding
interactions.  We have limited the types of interactions to include only
the minimal set required in the Standard Model and its supersymmetric
generalization. 

However, even with the minimal field content, there are several
other superpotential terms that one could envision adding to (\ref{WMSSM}) which
are consistent with all of the gauge symmetries of the theory.  Specifically, one could 
consider adding any or all of the following terms that violate $R$-parity:
\beq
W_{R} = {1\over 2} \lambda \epsilon_{ij} L^i L^j {e^c}
+ \lambda^{\prime} \epsilon_{ij} L^i Q^j {d^c} + {1\over 2} \lambda^{\prime\prime}
{u^c}{d^c}{d^c}
+ \mu^{\prime} L^i H_2^i .
\label{wr}
\eeq 
Each  of the terms in (\ref{wr}) has one or more suppressed generation indices.
We note that the terms proportional to $\lambda, \lambda^\prime$, and
$\mu^\prime$ both violate lepton number by one unit, whereas the term proportional
to $\lambda^{\prime\prime}$ violates baryon number by one unit. 

Each of the terms in (\ref{wr}) predicts new particle interactions and can be to some
extent constrained by the lack of observed exotic phenomena. In particular, any
combination of terms which violate both baryon and lepton number would be
unacceptable, unless the product of coefficients was extremely small. 
For example, consider the possibility that both $\lambda^\prime$
and $\lambda^{\prime\prime}$ were non-zero.  This would lead to the following
proton decay processes: $p \to e^+ \pi^0, \mu^+ \pi^0, \nu \pi^+,
\nu K^+$, etc.
The rate of proton decay due to this process would have no
suppression by any superheavy masses, since there is no GUT- or
Planck-scale physics involved: this is a purely (supersymmetric)
Standard Model interaction involving only the electroweak scale. 
The (inverse) rate can be easily estimated to be 
\beq
\Gamma^{-1}_p \sim {\tilde{m}^4 \over m_p^5} \sim 10^8 {\rm GeV}^{-1},
\eeq
assuming a supersymmetry breaking scale of $\tilde{m}$ of order 100 GeV. 
This should be compared with current limits to the proton life-time of $\gsim
10^{63}$ GeV$^{-1}$. Clearly the product of $\lambda^\prime$
and $\lambda^{\prime\prime}$ must be very small, if not exactly zero.

It is possible to eliminate the unwanted superpotential terms by imposing a
discrete symmetry on the theory called $R$-parity~\cite{Farrar:1978xj}.
This can be represented as
\beq
R = (-1)^{3B + L + 2s} ,
\label{Rparity}
\eeq
where $B,L$, and $s$ are the baryon number, lepton number, and spin
respectively. It is easy to see that, with the definition (\ref{Rparity}), all the known
Standard Model particles have $R$-parity +1. For example, the electron has
$B=0$, $L=-1$, and $s=1/2$, and the photon has $B=L=0$ and $s=1$, so in both cases
$R=1$. Similarly, it is clear that all superpartners of the known Standard
model particles have $R=-1$, since they must have the same value of $B$ and
$L$ as their conventional partners, but differ by 1/2 unit of spin. 
If $R$-parity is exactly conserved, then
all four superpotential terms in (\ref{wr}) must be absent. 

The additive conservation of the quantum numbers $B,L$, and $s$ implies
that $R$-parity must be conserved multiplicatively. A first important corollary 
is that the collisions of conventional particles must always produce supersymmetric 
particles in pairs, and a second corollary is that heavier supersymmetric
particles can decay only into lighter supersymmetric particles.
For our purposes here, an
even more important corollary of $R$-parity conservation is the prediction that the
lightest supersymmetric particle (LSP) must be stable, because it has no legal
decay mode. In much the same way that
baryon number conservation predicts proton stability, $R$-parity predicts
that the lightest $R = -1$ state is stable.  This makes supersymmetry an
extremely interesting theory from the astrophysical point of view, as the LSP
naturally becomes a viable dark matter candidate~\cite{Goldberg:1983nd,Ellis:1983ew}.  

\section{Possible Supersymmetric Dark Matter Candidates}

What options are available in the MSSM for the stable LSP?
Any electrically-charged LSP would bind to conventional matter,
and be detectable as an anomalous heavy nucleus,
since the `Bohr radius' for the LSP `atom' would be less than the nuclear radius.
Similarly, strongly-interacting LSPs would also form anomalous heavy nuclei.
However, experiments searching for such objects \cite{Rich:1987jd,Smith:1988ni,Hemmick:1989ns}
have excluded their presence on Earth
down to an abundance far lower than the expected abundance for the LSP (see below
for more details how this is calculated). Therefore, the stable LSP is presumably
electrically-neutral and can have only weak interactions. For this reason, the
commonly-expected signature of supersymmetric particle production at
colliders is missing energy carried away by undetected LSPs.

This still leaves us with several possible dark matter candidates in the MSSM,
specifically the sneutrino with spin zero, the neutralino with spin 1/2, and the gravitino
with spin 3/2. However, a sneutrino LSP would have relatively large coherent interactions
with heavy nuclei, and experiments searching directly for the scattering of 
massive dark matter particles on nuclei exclude a stable sneutrino weighing
between a few GeV and several TeV \cite{Falk:1994es}. The possible loophole of a very light
sneutrino was excluded by measurements of the invisible $Z$-boson decay rate at LEP
\cite{LEP:2003ih}.

The LSP candidate that is considered most often is the lightest neutralino.
In the MSSM there are four neutralinos, each of which is a  
linear combination of the following $R=-1$ neutral fermions~\cite{Ellis:1983ew}: the 
neutral wino $\tilde W^3$ (the partner of the third component of the SU(2)$_L$ 
triplet of weak gauge bosons); the U(1) bino $\tilde B$;
and two neutral Higgsinos $\tilde H_1$ and $\tilde H_2$ (the supersymmetric
partners of the neutral components of the two Higgs doublets).

The composition of the LSP $\chi$ can  be expressed as a linear combination
of these fields:
\begin{equation}
	\chi = \alpha \tilde B + \beta \tilde W^3 + \gamma \tilde H_1 +
\delta
\tilde H_2 ,
\label{chimix}
\end{equation}
whose mass and composition are determined by the SU(2)$_L$
and U(1) gaugino masses, $M_{2,1}$, the Higgs mixing parameter
$\mu$, and  $\tan \beta$, the ratio of the vacuum expectation values 
$v_{1,2} \equiv <0|H_{1,2}|0>$ of the two
neutral Higgs fields $\tan \beta \equiv v_2/v_1$. The mass of the LSP $\chi$
and the mixing coefficients $\alpha, \beta, \gamma$ and $\delta$ in (\ref{chimix})
for the neutralino components that compose the LSP 
can be found by diagonalizing the mass matrix
\begin{equation}
      ({\tilde W}^3, {\tilde B}, {{\tilde H}^0}_1,{{\tilde H}^0}_2 )
  \left( \begin{array}{cccc}
M_2 & 0 & {-g_2 v_1 \over \sqrt{2}} &  {g_2 v_2 \over \sqrt{2}} \\
0 & M_1 & {g_1 v_1 \over \sqrt{2}} & {-g_1 v_2 \over \sqrt{2}} \\
{-g_2 v_1 \over \sqrt{2}} & {g_1 v_1 \over \sqrt{2}} & 0 & -\mu \\
{g_2 v_2 \over \sqrt{2}} & {-g_1 v_2 \over \sqrt{2}} & -\mu & 0 
\end{array} \right) \left( \begin{array}{c} {\tilde W}^3 \\
{\tilde B} \\ {{\tilde H}^0}_1 \\ {{\tilde H}^0}_2 \end{array} \right),
\end{equation}
In different regions of the supersymmetric parameter space, the LSP
may be more bino-like, wino-like, or Higgsino-like, depending
on the relative magnitudes of the coefficients $\alpha, \beta, \gamma$ and $\delta$.
 
The relic abundance of an LSP candidate such as the lightest neutralino is 
calculated by solving the Boltzmann
equation for the LSP number density in an expanding Universe:
 \beq
{dn \over dt} = -3{{\dot R} \over R} n - \langle \sigma v \rangle (n^2 -
n_0^2) ,
\label{rate}
\eeq
where $n_0$ is the equilibrium number density of neutralinos.
Defining the quantity $f \equiv n/T^3$, we can rewrite this equation in terms of 
the reduced temperature $x \equiv T/m_\chi$:
\beq
{df \over dx} = m_\chi \left( {8 \pi^3 \over 90}G_N N \right)^{-1/2} \langle \sigma v \rangle
(f^2 - f_0^2) ,
\label{rate2}
\eeq
where $G_N$ is Newton's constant and $N$ is the number of relativistic degrees of 
freedom at a given temperature.
The solution to this equation at late times and low temperatures,
and hence small $x$, yields a constant value of
$f$, so that $n \propto T^3$. 

The technique \cite{Srednicki:1988ce} used to determine the neutralino
relic density is similar to that used previously for computing
the relic abundance of massive neutrinos~\cite{Hut:1977zn,Lee:1977ua,Vysotsky:1977pe},
with the substitution of  the appropriate annihilation cross section.
This and hence the relic density depend on additional parameters in the MSSM beyond $M_1, M_2,
\mu$, and $\tan \beta$, which include the sfermion masses, $m_{\tilde f}$ and mass of the
pseudoscalar Higgs boson, $m_A$.  In much of the parameter
space of interest, the LSP is a bino and the annihilation proceeds mainly
through crossed $t$-channel sfermion exchange. The exception is if
the sum of two neutralino masses happens to lie near a direct-channel
pole, such as $m_\chi \simeq$ $m_Z/2$ or $m_h/2$, 
in which case there are large contributions to the
annihilation through direct $s$-channel resonance exchange.
Since the neutralino is a Majorana fermion, away from such a resonance
the $s$-wave part of the annihilation cross section is generally
suppressed by the outgoing fermion masses, and the annihilation occurs mainly
through the $p$ wave, which is also suppressed because the annihilating LSPs are
non-relativistic at low temperatures (small $x$). This means that one can
approximate the annihilation cross section including $p$-wave corrections by
incorporating a term proportional to the temperature if neutralinos
are in thermal equilibrium: $\sigma v = a + b x + \dots $, where the
expansion coefficients $a, b$ are model-dependent.

Annihilations in the early Universe continue until the annihilation rate $\Gamma
\simeq \sigma v n_\chi$ drops below the expansion rate, after
which it is a good first approximation to assume that annihilations are negligible -
the freeze-out approximation. 
The final neutralino relic density, expressed as a fraction $\Omega_\chi$
of the critical energy density and denoting the present-day Hubble expansion rate
as $h$ in units of 100~km/s/Mpc, can be written as~\cite{Ellis:1983ew}
\begin{equation}
\Omega_\chi h^2 \simeq 1.9 \times 10^{-11} \left({T_\chi \over
T_\gamma}\right)^3 N_f^{1/2} \left({{\rm GeV} \over ax_f + {1\over 2} b
x_f^2}\right) ,
\label{relic}
\end{equation} 
where $(T_\chi/T_\gamma)^3$ accounts for the subsequent reheating of the
photon temperature with respect to $\chi$, due to the annihilations of
particles with mass $m < x_f m_\chi$~\cite{Steigman:1979xp,Olive:1980wz},
and $x_f = T_f/m_\chi$ is
proportional to the freeze-out temperature $T_f$. Eq. (\ref{relic} ) yields
a very good approximation to the relic density
except near direct $s$-channel annihilation poles or
thresholds, and in regions where the LSP is
nearly degenerate with the next lightest supersymmetric particle~\cite{Griest:1990kh}.

When there are several
particle species $i$ that are nearly degenerate in mass,
coannihilations between the different species become important. 
In this case~\cite{Griest:1990kh}, the rate equation
(\ref{rate}) still applies, provided $n$ is
interpreted as the total number density, 
\beq
n \equiv \sum_i n_i \;,
\label{n}
\eeq
$n_0$ is interpreted as the total equilibrium number density, 
\beq
n_0 \equiv  \sum_i n_{0,i} \;,
\label{neq}
\eeq
and the effective annihilation cross section as
\beq
\langle\sigma_{\rm eff} v_{\rm rel}\rangle \equiv
\sum_{ij}{ n_{0,i} n_{0,j} \over n_0^2}
\langle\sigma_{ij} v_{\rm rel}\rangle \;.
\label{sv2}
\eeq
In eq.~(\ref{rate2}),  $m_\chi$ is now understood to be the mass of the
lightest sparticle under consideration.

We turn finally to the third LSP candidate within the MSSM, namely the
gravitino. Since it has only gravitational-strength interactions, it is not
expected to have been in thermal equilibrium in the early Universe.
However, it could have been produced in high-energy particle
collisions in the early Universe, or in the decays of heavier
supersymmetric particles. The fact that the gravitino has only
gravitational-strength interactions implies that only decays of the
next-to-lightest supersymmetric particle (NLSP) would be significant
sources of gravitinos, and the NLSP would be metastable. As we
discuss in more detail later, there are important cosmological and
astrophysical constraints on the possible mass and lifetime of the
NLSP, derived principally from the agreement between
astrophysical observations and Big-Bang Nucleosynthesis
calculations of light-element abundances.

What might be the nature of the NLSP
be in such a gravitino LSP scenario? One option is the lighter of the two
supersymmetric partners of the $\tau$ lepton, denoted by $\tilde \tau_1$.
Being a metastable charged particle, it would have a distinctive experimental
signature at the LHC or other colliders. Studies within such a scenario
have shown that the mass of the ${\tilde \tau_1}$
could be measured very accurately, and that one could easily
reconstruct heavier sparticles that decay into the ${\tilde \tau_1}$~\cite{Ellis:2006vu}.

Alternatively, the NLSP might be the lighter supersymmetric partner
of the top quark, denoted by ${\tilde t_1}$~\cite{DiazCruz:2007fc,Santoso:2007uw,Kohri:2008cf},
which would have even more distinctive signatures at the LHC.
Immediately after production, it would become confined
inside a charged or neutral hadron. As it moves through an LHC
detector, it would have a high probability of changing its charge
as it interacts with the material in the detector. This combined with
its non-relativistic velocity would provide a truly distinctive
signature.

Yet another possibility is that the NLSP might be some flavour of
sneutrino~\cite{Ellis:2008as}, in which case the characteristic signature would be
missing energy carried away by the metastable sneutrino. This
could nevertheless be distinguished from the conventional
missing-energy signature of a neutralino LSP (or NLSP),
because the final states would be more likely to include the
charged lepton with the same flavour as the sneutrino NLSP, 
either $e$, $\mu$ or $\tau$.

These are just a few examples of the possible alternatives to
the conventional missing-energy signature of supersymmetry.
Studies have shown that the LHC would also have good prospects
for detecting such signatures.

\section{Renormalization-Group Equations and Electroweak
Symmetry Breaking}

The fact that measurements of the strengths of the Standard Model
gauge interactions measured at low energies are in excellent agreement
with the predictions of a supersymmetric gauge 
theory~\cite{Ellis:1990zq,Ellis:1990wk,Amaldi:1991cn,Langacker:1991an,Giunti:1991ta}  
was already cited as an important motivation for low-energy supersymmetry. It can also be
regarded as a motivation for thinking that other parameters of the
effective low-energy theory, e.g., the soft supersymmetry breaking parameters
can also be calculated and related using renormalization-group equations (RGEs)
below the grand unification scale.
For example, the one-loop RGEs for the gaugino masses are:
\beq
{dM_i \over dt} = -b_i \alpha_i M_i /4 \pi
\label{rengp}
\eeq
If  the gaugino masses have a common value $m_{1/2}$ at the grand unification scale,
these equations can be used to relate the physical low-energy, on-shell values of the 
gaugino masses to the corresponding gauge coupling strengths $\alpha_i$:
\beq
M_i (t)  = {\alpha_i (t) \over \alpha_i(M_{GUT})} m_{1/2} ,
\label{gauginorge}
\eeq
which implies that 
\beq
{M_1 \over g_1^2}={M_2 \over g_2^2}={M_3 \over g_3^2} 
\label{mgaugino}
\eeq
at the one-loop level.
When applying this relation within a specific grand unified theory, one must remember to
incorporate the difference of the normalization of the U(1) factor from that
in the Standard Model, so that we have $M_1 = {5\over 3} {\alpha_1\over \alpha_2} M_2$
for the one-loop relation between the bino and wino masses. Also, the simple relations
(\ref{mgaugino}) are modified by threshold corrections at the electroweak scale, and by
two-loop effects in the RGEs.
The soft supersymmetry-breaking scalar masses-squared $m^2$ and the trilinear
couplings $A$ are renormalized analogously to (\ref{rengp}), with the difference that
Yukawa interactions contribute as well as gauge interactions. However, the Yukawa
contributions are small, except for the supersymmetric partners of third-generation
fermions.

As described above, the MSSM has over 100 undetermined parameters, which are
mainly associated with the breaking of supersymmetry. It is often assumed that the
soft supersymmetry-breaking parameters $M^a, m^2$ and $A$ have some
universality properties. There are phenomenological arguments, based on the
success of the Standard Model in describing 
the observed suppression of flavour-changing interactions, that,
at some input scale (often assumed to be that of grand unification),
the parameters $m^2$ and $A$ must be universal for supersymmetric particles with the same
gauge quantum numbers, e.g., the supersymmetric partners of the $e, \mu$ and $\tau$.
There is no strong argument why these parameters should be universal for
supersymmetric particles with different quantum numbers, e.g., $d, u$ and $e$,
though this may occur in some grand unified theories, as may unification of the
gaugino masses $M^a$. The simplified version of the MSSM in which universality
at the grand unification scale is assumed for each of  $M^a, m^2$ and $A$ is
called the constrained MSSM (CMSSM)\cite{Drees:1992am,Baer:1995nc,Baer:1997ai,Baer:2000jj,Lahanas:2000xd,Ellis:1996xu,Ellis:1997wva,Ellis:1998jk,Barger:1997kb,Ellis:2000we,Barger:2001yy,Roszkowski:2001sb,Lahanas:2001yr,Djouadi:2001yk,Chattopadhyay:2001va,Ellis:2002rp,Baer:2002gm,Arnowitt:2002he,
Ellis:2001msa,Ellis:2003cw,Baer:2003yh,Lahanas:2003yz,Chattopadhyay:2003xi,Munoz:2003gx}.

Once one has chosen a set of boundary conditions at the grand unification scale
and run the RGEs down to the electroweak scale, one must check the properties
of the electroweak vacuum, which are characterized by specifying the mass of the
$Z$ boson, $M_Z$, and the ratio of the two Higgs vacuum expectation values, 
$\tan \beta$. These electroweak symmetry-breaking conditions should be used as
consistency conditions on the solutions to the RGEs, e..g., of the CMSSM. They are
frequently used to fix, as functions of the input values of the common gaugino mass
$m_{1/2}$, $m$, $A$ and $\tan \beta$, 
the magnitudes of the Higgs mixing mass parameter, $\mu$, and
of the bilinear coupling, $B$, which determines the pseudoscalar Higgs mass, $m_A$.
The sign of $\mu$ remains free.

\begin{figure}[ht]
\begin{center}
\resizebox{0.6\textwidth}{!}{%
  \includegraphics{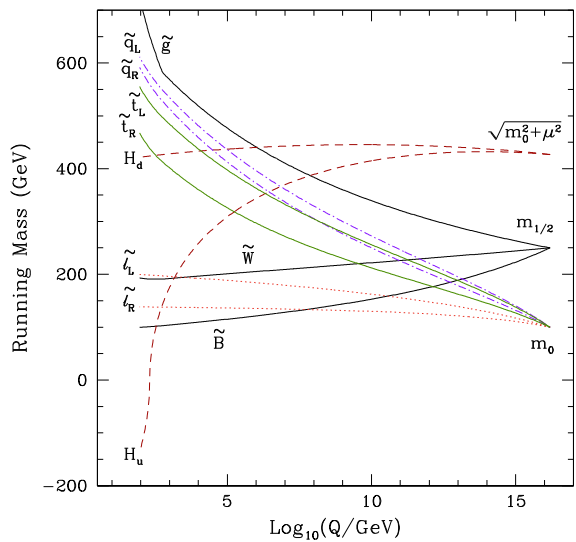}
}
\end{center}
\caption{\it The renormalization-group evolution of the mass parameters in the CMSSM,
assuming $m_{1/2} = 250$~GeV, $m_0 = 100$~GeV, $\tan \beta = 3$, $A_0 = 0$, and $\mu < 0$.
We thank Toby Falk for providing this figure.}
\label{running}       
\end{figure}

An example of the running of the mass parameters in the CMSSM as functions
of the renormalization scale is shown
in Fig.~\ref{running}, using as inputs the choices $m_{1/2} = 250$~GeV, $m_0 = 100$~GeV, 
$\tan \beta = 3$, $A_0 = 0$, and $\mu < 0$.
We notice in the figure several characteristic features of the sparticle spectrum.
For example, the colored sparticles are typically the heaviest, 
because of the large positive corrections to their masses arising from 
$\alpha_3$-dependent terms in the RGEs.  Also, one finds that the bino,
$\widetilde{B}$, is typically the lightest
sparticle.  Most importantly, we notice that one of the Higgs masses squared, goes
negative, triggering electroweak symmetry
breaking~\cite{Ibanez:1982fr,Ibanez:1982ee,Ellis:1982wr,Ellis:1983bp,AlvarezGaume:1983gj}. 
(The negative
sign in the figure refers to the sign of the mass squared, even though it is the
mass of the sparticles which is depicted.) 

\section{The CMSSM}

\begin{figure}[ht]
\resizebox{0.45\textwidth}{!}{
\includegraphics{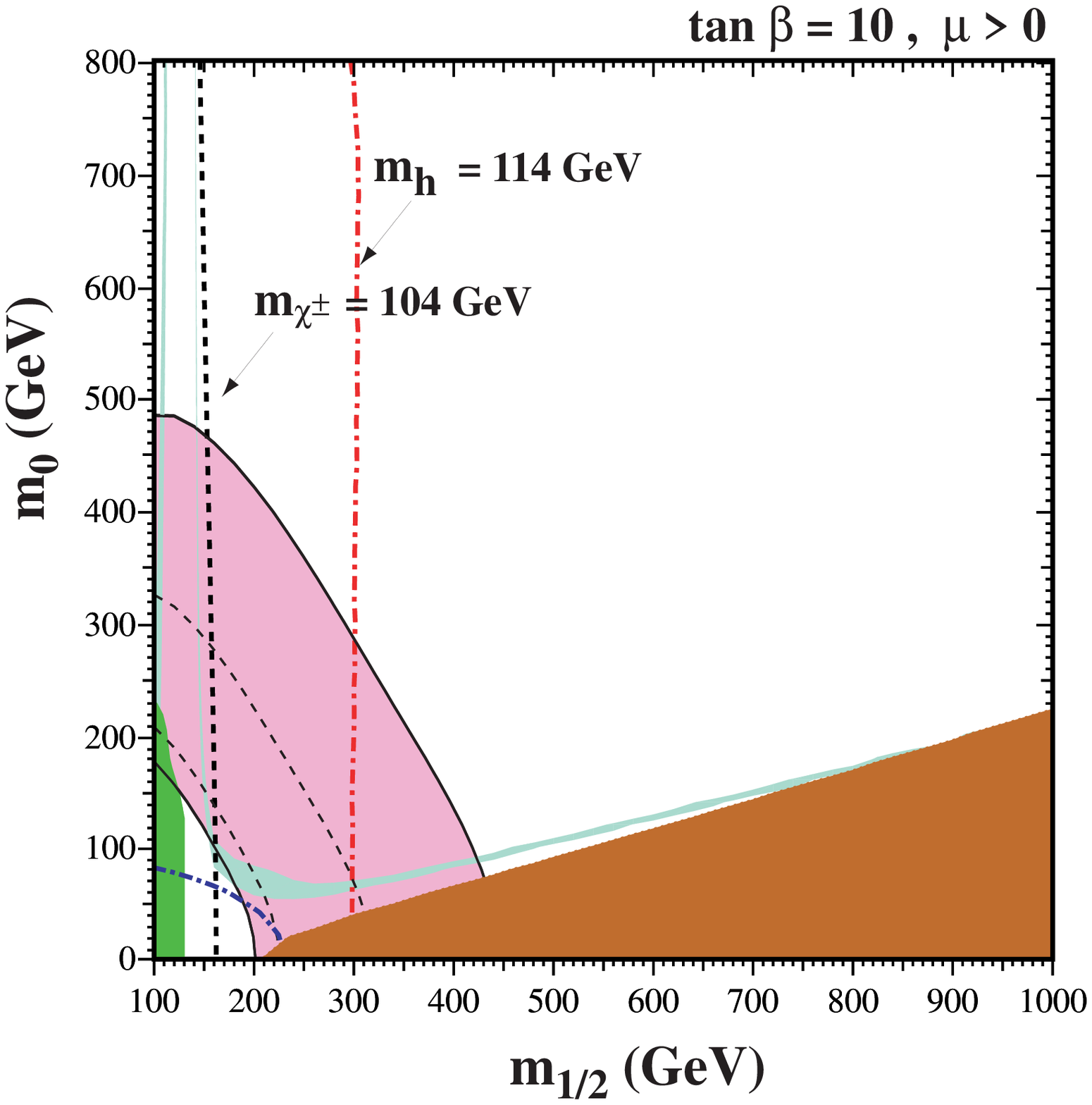}}
\resizebox{0.45\textwidth}{!}{
\includegraphics{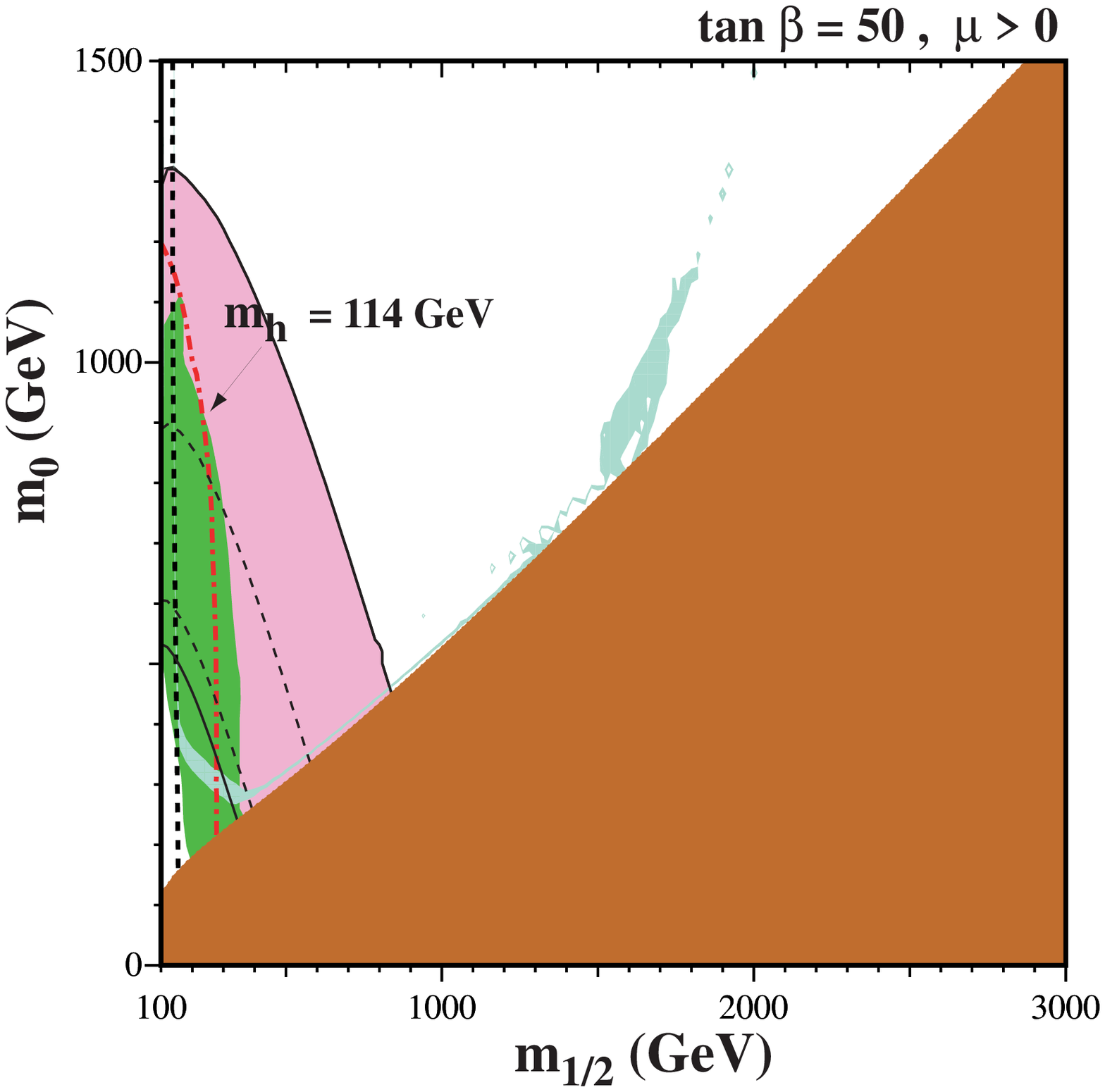}}
\caption{\label{fig:UHM}
{\it The $(m_{1/2}, m_0)$ planes for  (a) $\tan \beta = 10$ and (b) $\tan \beta = 50$,
assuming $\mu > 0$, $A_0 = 0$, $m_t = 175$~GeV and
$m_b(m_b)^{\overline {MS}}_{SM} = 4.25$~GeV. The near-vertical (red)
dot-dashed lines are the contours for $m_h = 114$~GeV, and the near-vertical (black) dashed
line is the contour $m_{\chi^\pm} = 104$~GeV. Also
shown by the dot-dashed curve in the lower left is the region
excluded by the LEP bound $m_{\tilde e} > 99$ GeV. The medium (dark
green) shaded region is excluded by $b \to s
\gamma$, and the light (turquoise) shaded area is the cosmologically
preferred region. In the dark
(brick red) shaded region, the LSP is the charged ${\tilde \tau}_1$. The
region allowed by the E821 measurement of $a_\mu$ at the 2-$\sigma$
level, is shaded (pink) and bounded by solid black lines, with dashed
lines indicating the 1-$\sigma$ ranges.}}
\end{figure}

For given values of $\tan \beta$, $A_0$,  and $sgn(\mu)$, the regions of the CMSSM
parameter space that yield an
acceptable relic density and satisfy the other phenomenological constraints
may conveniently be displayed in the  $(m_{1/2}, m_0)$ plane.
Fig.~\ref{fig:UHM} displays, for $\tan \beta = 10$ (a) and 50 (b),
the impacts of the most relevant constraints.
These include the LEP lower limits on the chargino mass: $m_{\chi^\pm} > 104$~GeV~\cite{LEPsusy}, 
on the selectron mass: $m_{\tilde e} > 99$~GeV~ \cite{LEPSUSYWG_0101} 
and on the Higgs mass: $m_h >
114$~GeV~\cite{Barate:2003sz,LEPHiggs}. The former two constrain $m_{1/2}$ and $m_0$ directly
via the sparticle masses, and the latter indirectly via the sensitivity of
radiative corrections to the Higgs mass to the sparticle masses,
principally $m_{\tilde t, \tilde b}$. Here the code
{FeynHiggs}~\cite{Heinemeyer:1998yj,Heinemeyer:1998np} is used for the calculation of $m_h$.
It would be prudent to assign an uncertainty of 3~GeV to this calculation. 
Nevertheless, the Higgs limit  imposes important constraints,
principally on $m_{1/2}$ and particularly at low $\tan \beta$.
Another constraint is the requirement that
the branching ratio for $b \rightarrow
s \gamma$ be consistent with the experimental measurements~\cite{Barberio:2006bi}. 
These measurements agree with the Standard Model, and
therefore provide bounds on MSSM particles~\cite{Degrassi:2000qf},  such as the chargino and
charged Higgs bosons, in particular. Typically, the $b\rightarrow s\gamma$
constraint is more important for $\mu < 0$, but it is also relevant for
$\mu > 0$,  particularly when $\tan\beta$ is large. The constraint imposed by
measurements of $b\rightarrow s\gamma$ also exclude small
values of $m_{1/2}$. Finally, there are
regions of the $(m_{1/2}, m_0)$ plane that are favoured by
the Brookhaven National Laboratory measurement~\cite{Bennett:2006fi} of $g_\mu - 2$. 
Here we assume
the Standard Model calculation~\cite{Passera:2008jk} of $g_\mu - 2$
using $e^+ e^-$ data, and indicate by dashed and solid lines the contours
of 1- and 2-$\sigma$ level deviations induced by supersymmetry.  

The most precise constraint on supersymmetry may be that provided by the density
of cold dark matter, as determined from astrophysical and cosmological measurements
by WMAP and other experiments~\cite{Dunkley:2008ie}:
\begin{equation}
\Omega_{CDM} \; = \; 0.1099 \pm 0.0062 .
\label{OCDM}
\end{equation}
Applied straightforwardly to the relic LSP density $\Omega_{LSP} h^2$,
this would give a very tight relation between supersymmetric
model parameters, fixing some combination of them at the \% level,
which would essentially reduce the dimensionality of the supersymmetric 
parameter space by one unit. Let us
assume for now that the LSP is the lightest neutralino $\chi$, whose density is
usually thought to be fixed by freeze-out from thermal equilibrium in the
early Universe, as discussed previously.
In this case, respecting the constraint (\ref{OCDM}) would force the CMSSM
into one of the narrow WMAP `strips' in planar projections of the parameters~\cite{Ellis:2003cw},
as illustrated by the narrow light (turquoise) regions in Fig.~\ref{fig:UHM}. However,
caution should be exercised before jumping to this conclusion.

Supersymmetry might not be the only contribution to the cold dark matter,
in which case (\ref{OCDM}) should be interpreted as an upper limit on $\Omega_{LSP} h^2$.
However, most of the supersymmetric parameter
space in simple models gives a supersymmetric relic density that exceeds the WMAP range
(\ref{OCDM}), e.g., above the WMAP `strip' in Fig.~\ref{fig:UHM},
and the regions with lower density generally correspond to
{\it lower} values of the sparticle masses, i.e., below
the WMAP `strip' in Fig.~\ref{fig:UHM}.

However, even if one takes them seriously, the locations of these WMAP `strips' do vary 
significantly with the choices of other supersymmetric parameters, as can be seen by
comparing the cases of $\tan \beta = 10, 50$ in Fig.~\ref{fig:UHM}(a, b). 
As one varies $\tan \beta$, the WMAP `strips' cover much of the $(m_{1/2}, m_0)$ plane.

Several different regions of the WMAP `strips' in the
CMSSM $(m_{1/2}, m_0)$ plane can be distinguished, in
which different dynamical processes are dominant. At low values of $m_{1/2}$
and $m_0$, simple $\chi - \chi$ annihilations via crossed-channel sfermion
exchange are dominant, but this `bulk' region is now largely excluded by the LEP
lower limit on the Higgs mass, $m_h$. At larger $m_{1/2}$, but relatively small $m_0$,
close to the boundary of the region where the lighter stau is lighter than the
lightest neutralino: $m_{\tilde \tau_1} < m_\chi$, coannihilation between the $\chi$
and sleptons are important in suppressing the relic $\chi$ density into the WMAP
range (\ref{OCDM}), as seen in Fig.~\ref{fig:UHM}. 
At larger $m_{1/2}, m_0$ and $\tan \beta$, the relic $\chi$
density may be reduced by rapid annihilation through direct-channel $H, A$ Higgs 
bosons, as seen in Fig.~\ref{fig:UHM}(b). Finally, the relic density can again be brought
down into the WMAP range (\ref{OCDM}) at large $m_0$ (not shown in 
Fig.~\ref{fig:UHM}), in the `focus-point' region close the boundary where electroweak 
symmetry breaking ceases to be possible and the lightest neutralino $\chi$
acquires a significant higgsino component \cite{Feng:1999zg}.

As seen in Fig.~\ref{fig:UHM}, the relic density constraint is compatible
with relatively large values of $m_{1/2}$ and $m_0$, and it is interesting
to look for any indication where the supersymmetric mass scale might lie
within this range, using the available phenomenological and cosmological constraints.
A global likelihood analysis enables one to 
pin down the available parameter space in the CMSSM and the related models
discussed later. One can avoid the dependence on priors by performing
a pure likelihood analysis as in~\cite{Ellis:2003si}, or a purely $\chi^2$-based fit as 
done in~\cite{Ellis:2007fu,Buchmueller:2007zk}.  
Here we present results from one such analysis~\cite{Buchmueller:2008qe}, 
which used a Markov-Chain
Monte Carlo (MCMC) technique to explore efficiently the likelihood function in
the parameter space of the CMSSM. A full list of the observables and the values assumed
for them in this global analysis are given in~\cite{Buchmueller:2007zk}, as updated 
in~\cite{Buchmueller:2008qe}. 

\begin{figure}[ht]
\begin{center}
\begin{picture}(300,200)
  \put( 20,   20){ \resizebox{0.65\textwidth}{!}{\includegraphics{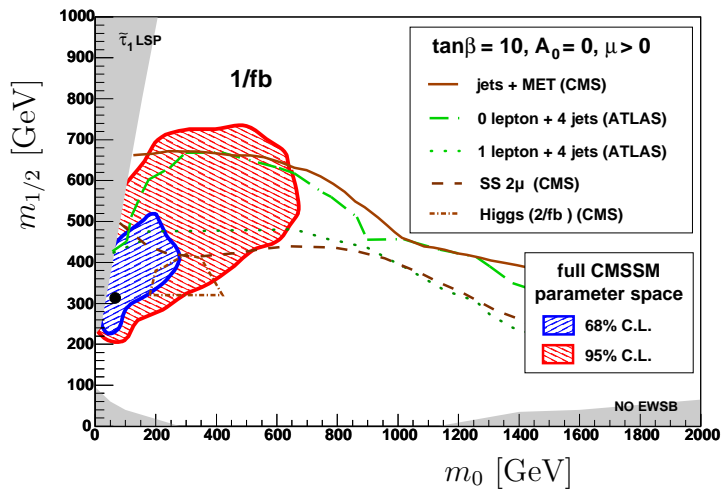}}}
  \put(170,   5){$m_0$ [GeV]}
  \put( 15,   100){\begin{rotate}{90}$m_{1/2}$ [GeV]\end{rotate}}
\end{picture}
\end{center}
\caption{\label{fig:MCMC}
{\it The $(m_0, m_{1/2})$ plane in the CMSSM
showing the regions favoured in a likelihood analysis
at the 68\% (blue) and 95\% (red) confidence levels~\protect\cite{Buchmueller:2008qe}. The best-fit
point is shown as the black point. Also shown are
the discovery contours in different channels for the LHC with 1/fb
(2/fb for the Higgs search in cascade decays of sparticles)~\protect\cite{ATLAS:1999fr,Ball:2007zza}.}}
\end{figure}

The 68\% and 95\% confidence-level (C.L.) regions in the
$(m_{1/2}, m_0)$ plane of the CMSSM are shown in
Fig.~\ref{fig:MCMC}~\cite{Buchmueller:2008qe}. Also shown for comparison are the physics
reaches of ATLAS and CMS with 1/fb of integrated luminosity~\cite{ATLAS:1999fr,Ball:2007zza}. 
(MET stands for missing
transverse energy, SS stands for same-sign dilepton pairs, and the
sensitivity for finding the lightest Higgs boson in cascade decays of
supersymmetric particles is calculated for 2/fb of data.) The likelihood
analysis assumed $\mu > 0$, as motivated by the sign of the
apparent discrepancy in $g_\mu - 2$, but sampled all values of $\tan \beta$
and $A_0$: the experimental sensitivities were estimated
assuming $\tan \beta = 10$ and $A_0 = 0$, but are probably not
very sensitive to these assumptions. The global maxima of the
likelihood function (indicated by the black dot) is at
$m_{1/2} = 310$~GeV,
$m_0 = 60$~GeV, $A_0 = 240$~GeV, $\tan \beta = 11$ and
$\chi^2/N_{dof} = 20.4/19$ (37\% probability). It is encouraging that the best-fit points lie
well within the LHC discovery range, as do the 68\% and most of the
95\% C.L. regions. It is also encouraging that the two best-fit points
have similar values of $m_{1/2}, m_0$ and $\tan \beta$, the most
important parameters for the sparticle spectrum, indicating that the 
likelihood analysis is relatively insensitive to the theoretical model
assumptions.

\begin{figure}[h]
\begin{center}
\resizebox{0.45\textwidth}{!}{
\includegraphics{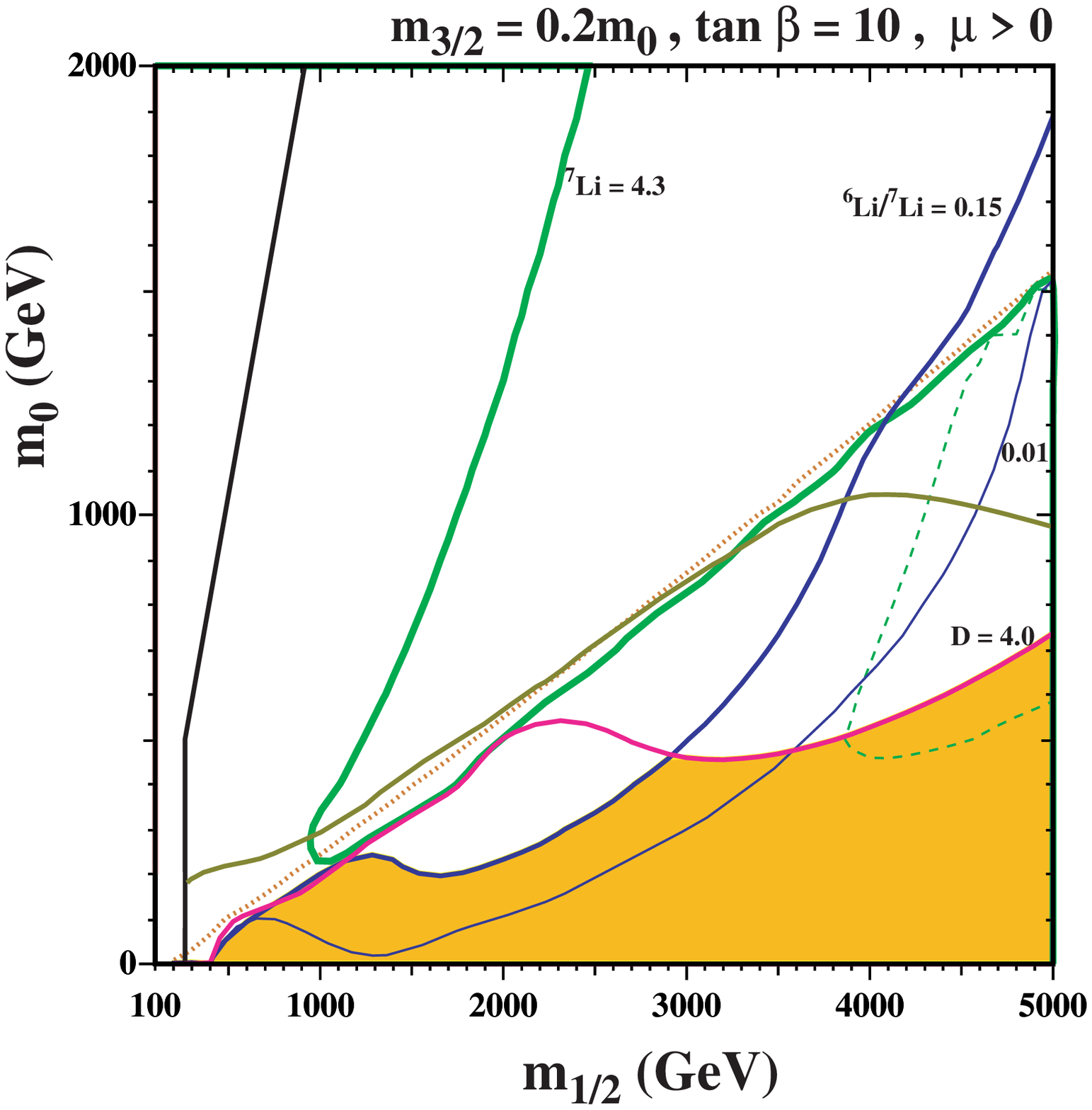}}
\resizebox{0.45\textwidth}{!}{
\includegraphics{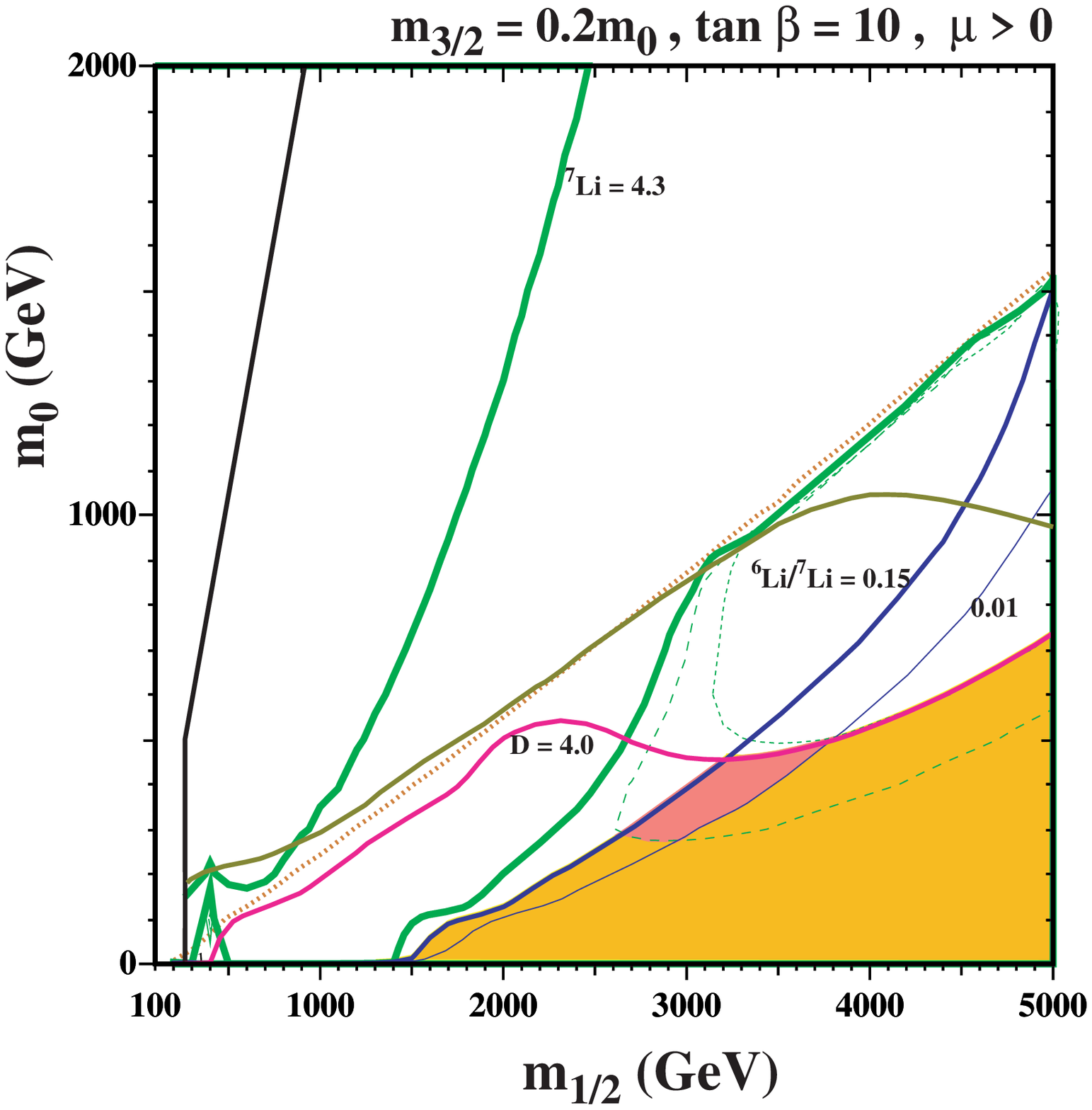}}
 \caption{\label{fig:cefos1}
{\it 
The $(m_{1/2}, m_0)$ planes for $m_t = 172.7$ GeV, $A_0=0$, $\mu > 0$ and $\tan
\beta = 10$ with $m_{3/2} = 0.2 m_0$ without  (a) and with (b) the effects of
metastable stau bound states included.
The regions to the left of the solid black lines are not considered, 
since there the gravitino is not the LSP.
In the orange (light) shaded regions, the differences between the calculated 
and observed light-element abundances are no greater than in standard 
BBN without late particle decays. In the pink (dark) shaded region in 
panel (b), the 
abundances lie within the ranges favoured by observation. 
The significances of the 
other  lines and contours  are explained in the text.}}
\end{center}
\end{figure}

In contrast to this neutralino LSP scenario,
the gravitino dark matter (GDM) scenario in the CMSSM is tightly constrained by the
astrophysical constraints on the cosmological abundances of light elements, as
seen in Fig.~\ref{fig:cefos1}~\cite{Cyburt:2006uv}. However, such a scenario might have some advantages,
e.g., by enabling the cosmological prediction for the abundance of 
$^7$Li~\cite{Cyburt:2008kw} to be
improved, as also shown in Fig.~\ref{fig:cefos1}(b).

Recently, new attention has been focussed on the regions in which a metastable stau
is the next-to-lightest sparticle (NSP) in a GDM scenario,
due to its ability to form bound states (primarily with $^4$He). 
When such bound states occur, they catalyze certain nuclear
reactions such as $^4$He(D, $\gamma$)$^6$Li, which is normally highly suppressed
due to the production of a low-energy $\gamma$, whereas the bound-state reaction is 
not~\cite{Pospelov:2006sc,Hamaguchi:2007mp,Bird:2007ge}. In Fig. \ref{fig:cefos1}(a),  
the $(m_{1/2}, m_0)$ plane is displayed
showing explicit element abundance contours~\cite{Cyburt:2006uv} 
when the gravitino mass is  $m_{3/2} = 0.2 m_0$ in the 
{\it absence} of stau bound-state effects. To the left of the solid black line 
the gravitino is not the LSP.
The diagonal red dotted line corresponds to the boundary between
a neutralino and stau NSP: above it, the neutralino is the NSP,
and below it, the stau is the NSP.  Very close to this boundary,
there is a diagonal brown solid line.  Above this line, the relic
density of gravitinos from NSP decay is too high, i.e.,
\beq
\frac{m_{3/2}}{m_{NSP}} \Omega_{NSP} h^2 > 0.12.
\eeq
Thus we should restrict our attention to the area below this line.

The very thick green line labelled \li7 = 4.3
corresponds to the contour where \li7/H = $4.3 \times 10^{-10}$, a value very
close to the standard BBN result for \li7/H. 
It forms a `Vee' shape, whose right edge runs along the neutralino-stau NSP border.
Below the Vee, the abundance of \li7 is smaller than the standard BBN result. 
However, for relatively small values of $m_{1/2}$, 
the \li7 abundance does not differ very much from the standard BBN result:
it is only when $m_{1/2} \gsim 3000$~GeV that \li7 begins to drop significantly.  
The stau lifetime drops with increasing $m_{1/2}$, and when 
$\tau \sim 1000$~s, at $m_{1/2} \sim 4000$ GeV, the \li7 abundance has been reduced 
to an observation-friendly value close to $2 \times 10^{-10}$ as reported 
in~\cite{Jedamzik:2004er,Jedamzik:2004ip}
and shown by the (unlabeled) thin dashed (green) contours.

The region where the
\li6/\li7 ratio lies between 0.01 and 0.15 forms a band which moves
from lower left to upper right.  As one can see in the orange
shading, there is a large region where the lithium isotopic ratio can be
made acceptable. However, if we restrict to D/H $< 4.0 \times 10^{-5}$, we
see that this ratio is interesting only when \li7 is at or slightly below
the standard BBN result.

Turning now to Fig.~\ref{fig:cefos1}(b), we show the analogous results when
the bound-state effects are included in the calculation.  The abundance
contours are identical to those in panel (a) above the
diagonal dotted line, where the NSP is a neutralino and bound states do
not form.  We also note that the bound-state effects on D and \he3 are
quite minimal, so that these element abundances are very similar to those
in Fig.~\ref{fig:cefos1}(a).  However, comparing panels (a) and (b), one sees
dramatic bound-state effects on the lithium abundances.  
Everywhere to the left of the solid blue line labeled 0.15 is excluded.  
In the stau NSP region, this means that $m_{1/2} \gsim 1500$~GeV.  
Moreover, in the stau region to the right of the \li6/\li7 = 0.15 contour,
the \li7 abundance drops below $9 \times 10^{-11}$ (as shown by the thin
green dotted curve).
In this case, not only do the bound-state effects
increase the \li6 abundance when $m_{1/2}$ is small (i.e., at relatively
long stau lifetimes), but they also decrease the \li7 abundance when the
lifetime of the stau is about 1500~s. Thus, at $(m_{1/2}, m_0) \simeq
(3200,400)$~GeV, we find that \li6/\li7 $\simeq 0.04$, \li7/H $\simeq 1.2
\times 10^{-10}$, and D/H $\simeq 3.8 \times 10^{-5}$.  Indeed, when
$m_{1/2}$ is between 3000-4000 GeV, the bound-state effects cut the \li7
abundance roughly in half. In the darker (pink) region, 
the lithium abundances match the observational plateau
values, with the properties $\li6/\li7 > 0.01$ and $0.9 \times10^{-10} <
\li7/\textrm{H} < 2.0 \times10^{-10}$. This example demonstrates that it is 
possible to resolve the \li6/\li7 by postulating GDM with a stau NSP.

\section{mSUGRA}

Minimal supergravity (mSUGRA) is often used as a basis for
phenomenological studies~\cite{Barbieri:1982eh,Nilles:1983ge,Brignole:1997dp}. 
The framework termed above the CMSSM is occasionally referred to as the mSUGRA.
However, models based strictly on minimal supergravity should
employ two additional constraints~\cite{Ellis:2003pz,Ellis:2004qe}.
One is a relation between the soft supersymmetry-breaking bilinear and
trilinear parameters: $B_0 = A_0 - m_0$, and the other is a relation between the
gravitino and input scalar masses: $m_{3/2} = m_0$.
In the simplest version of
mSUGRA~\cite{Polonyi:1977pj,Barbieri:1982eh,Nilles:1983ge,Brignole:1997dp}, 
where supersymmetry is
broken by a single field in a hidden sector, the
universal trilinear soft
supersymmetry-breaking terms  are $A_0 = (3 -
\sqrt{3}) m_{0}$ and bilinear
soft supersymmetry-breaking term is $B_0 = (2 - \sqrt{3}) m_{0}$, which is a
special case of the general relation $B_0 = A_0 - m_0$. 

Given such a relation between $B_0$ and $A_0$, one can no longer use the
standard CMSSM boundary conditions, in which $m_{1/2}$, $m_0$, $A_0$,
$\tan \beta$, and $sgn(\mu)$ are input at the GUT scale, and then $\mu$ and $B$ 
are determined by the electroweak symmetry-breaking conditions.
In this case, it is natural to use $B_0$ as an input and calculate $\tan \beta$ 
from the minimization of the Higgs potential~\cite{Ellis:2003pz,Ellis:2004qe}.

\begin{figure}[h]
\begin{center}
\resizebox{0.45\textwidth}{!}{
\includegraphics{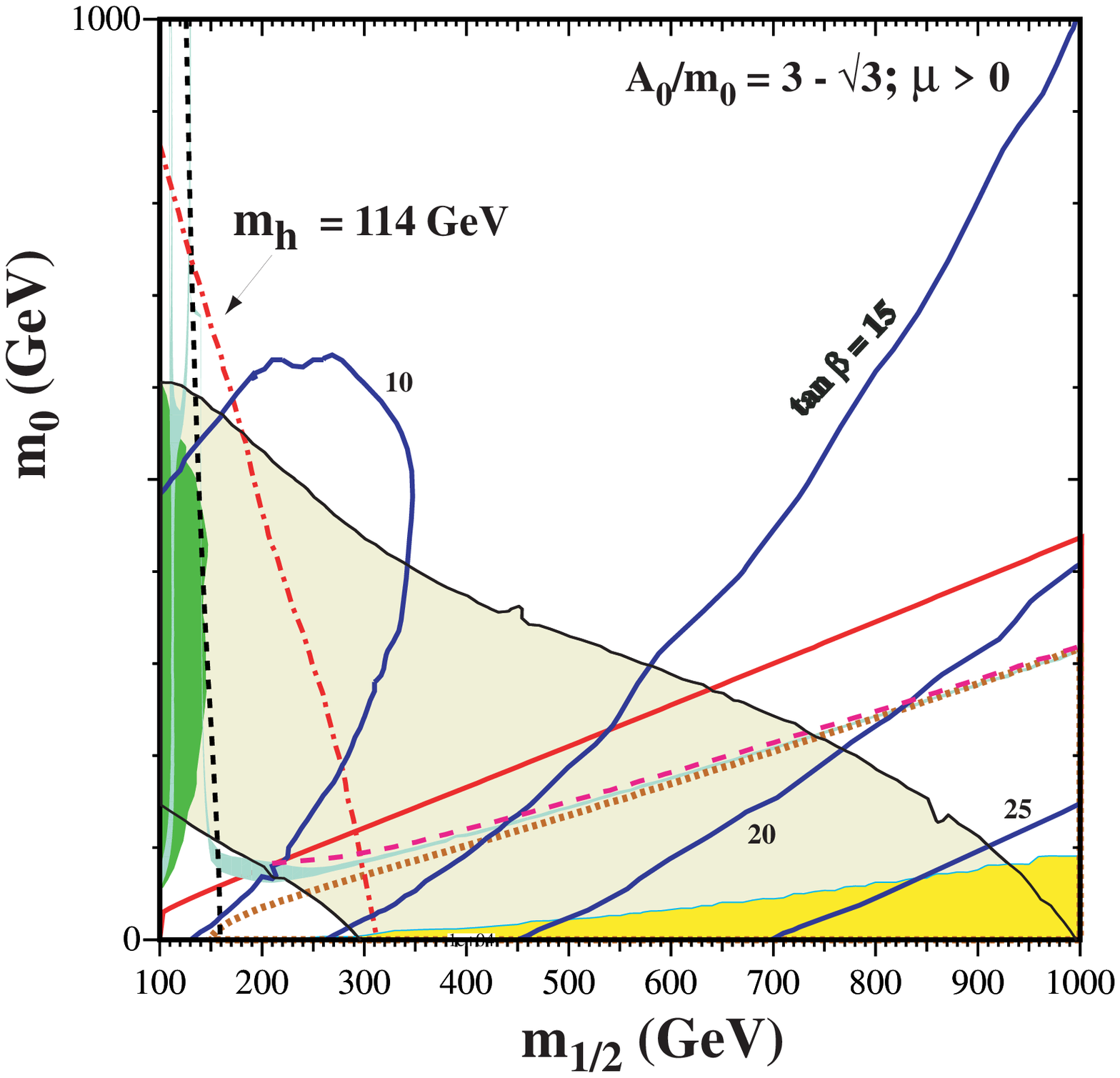}}
\resizebox{0.45\textwidth}{!}{
\includegraphics{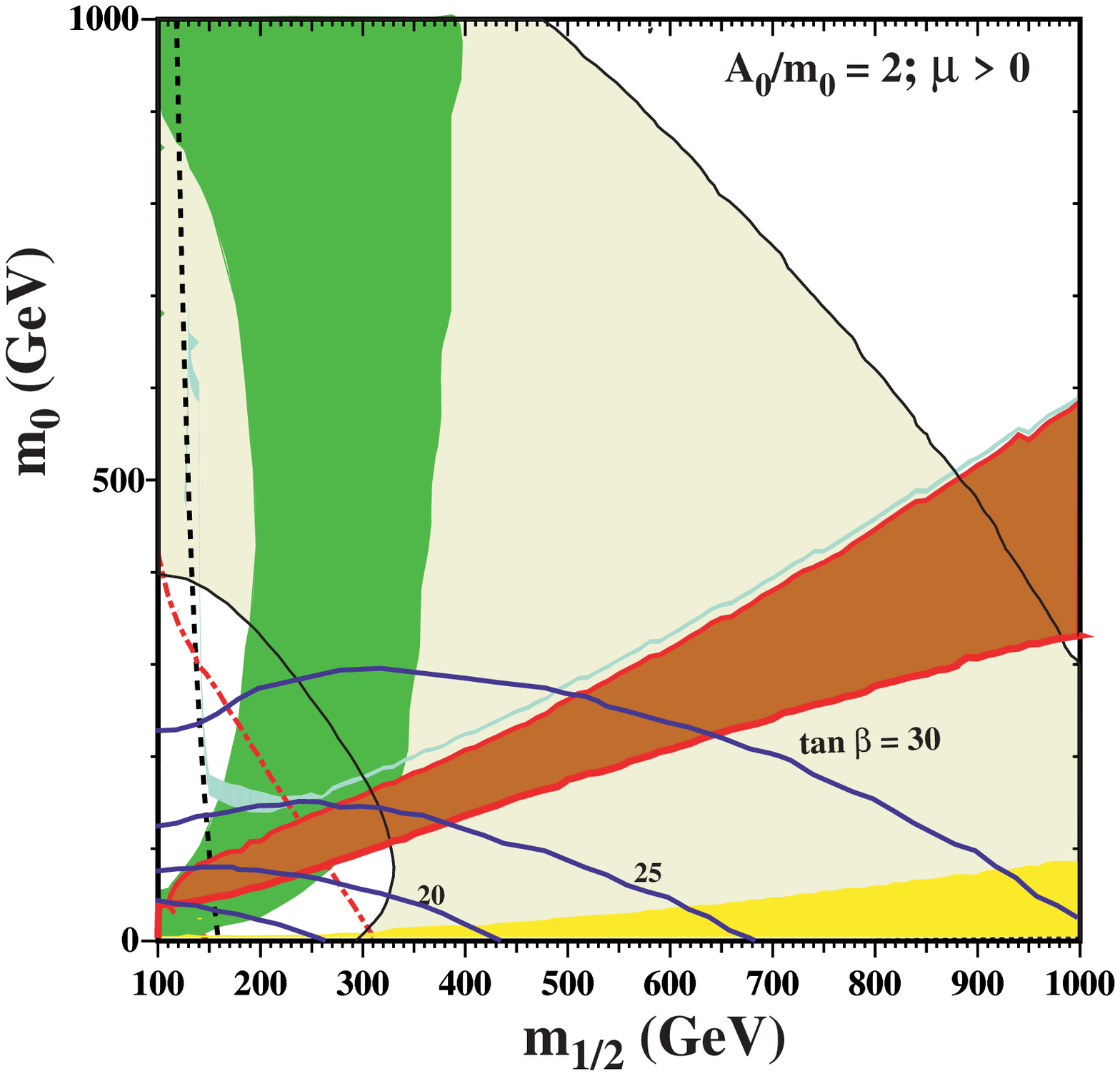}}
\caption{\label{fig:msugra}
{\it 
Examples of mSUGRA $(m_{1/2}, m_0)$ planes with contours of $\tan \beta$ 
superposed, for $\mu > 0$ and (a) the simplest Polonyi model with $A_0/m_0 = 3 - 
\sqrt{3}$,  and (b) $A_0/m_0 = 2.0$, all with $ B_0 =
A_0 -m_0$. In each panel, we show the regions excluded by 
the LEP lower limits on MSSM particles and those ruled out by $b
\to s \gamma$ decay (medium green shading): the regions 
favoured by $g_\mu - 2$ are very light (yellow) shaded, bordered by a thin
(black) line.
The dark (chocolate) solid lines separate the 
neutralino and gravitino LSP regions. The regions favoured 
by WMAP in the neutralino LSP case have light (turquoise)
shading. The dashed (pink) line corresponds to the maximum relic density 
for the gravitino LSP, and regions allowed by BBN constraint neglecting the effects of bound states on NSP 
decay are light (yellow) shaded.}}
\end{center}
\end{figure}

Phenomenologically distinct planes may be determined by specifying
a choice for $A_0/m_0$, with the above-mentioned simplest hidden sector
being one example.  In Fig. \ref{fig:msugra}, two such planes are shown assuming $m_t = 172.7$ 
GeV
with (a) $A_0/m_0 = 3 - \sqrt{3}$, as predicted in the simplest model of
supersymmetry breaking~\cite{Polonyi:1977pj}, and with (b) $A_0/m_0 = 2.0$.
We show in these $(m_{1/2}, m_0)$ planes the contours of $\tan \beta$ as solid
blue lines. Also shown are the contours where
$m_{\chi^\pm} > 104$~GeV (near-vertical black dashed lines) and $m_h >
114$~GeV (diagonal red dash-dotted lines). The regions excluded by $b \to
s \gamma$ have medium (green) shading, those where the relic density
of neutralinos lies within the WMAP range have light (turquoise) shading,
and the region suggested by $g_\mu - 2$ at 2-$\sigma$ has very light (yellow) shading,
as in the CMSSM planes shown previously.
As one can see, relatively low values of $\tb$ are obtained in most of the visible planes.

Another difference between the CMSSM and models based on
mSUGRA concerns the mass of the gravitino.
In the CMSSM, it is not specified and and can be taken suitably large so that the 
neutralino or the lighter stau is the LSP. In mSUGRA, the scalar masses
at the GUT scale,
$m_0$, are determined by (and equal to) the gravitino mass. In Fig. \ref{fig:msugra}, 
the gravitino LSP and the neutralino LSP regions are separated by dark (chocolate)
solid lines.
Above these lines, the neutralino (or stau) is the LSP, whilst below them the gravitino is the 
LSP~\cite{Feng:2003xh,Ellis:2003dn,Feng:2004mt}.
As one can see by comparing the two panels, the potential for neutralino dark matter
in mSUGRA models is dependent on $A_0/m_0$.  In panel (a), the only areas 
where the neutralino density are not too large occur where the Higgs mass
is far too small or, at higher $m_0$, the chargino mass is too small. 
At larger $A_0/m_0$, the coannihilation strip rises above the
neutralino-gravitino LSP boundary.  In panel (b), we see the familiar coannihilation strip.
It should be noted that the focus-point region is not realized in mSUGRA models
as the value of $\mu$ does not decrease with increasing $m_0$ when $A_0/m_0$ is 
fixed and $B_0 = A_0 - m_0$. There are also no funnel regions, as $\tb$ is never sufficiently high.

In the gravitino LSP regions, the NSP
may be either the neutralino or stau, which are now unstable.
(Note that in panel (b) there is also a region which is excluded because the stau is the LSP.)
The relic density of gravitinos is acceptably low only below the
dashed (pink) line. This excludes a supplementary domain of the $(m_{1/2},
m_0)$ plane in panel (a) which has a neutralino NSP (the dotted (red) curve in panel (a)
separates the neutralino and stau NSP regions). 
However, the strongest constraint is provided by the effect of neutralino or stau
decays on Big-Bang 
Nucleosynthesis~\cite{Cyburt:2002uv,Feng:2004zu,Ellis:2005ii,Kohri:2005wn,Cerdeno:2005eu,Steffen:2006hw}.  
Outside the light (yellow) shaded region,
the decays spoil the success of BBN.

\section{Other Possibilities}

\begin{figure}
\begin{center}
\resizebox{0.5\textwidth}{!}{%
  \includegraphics{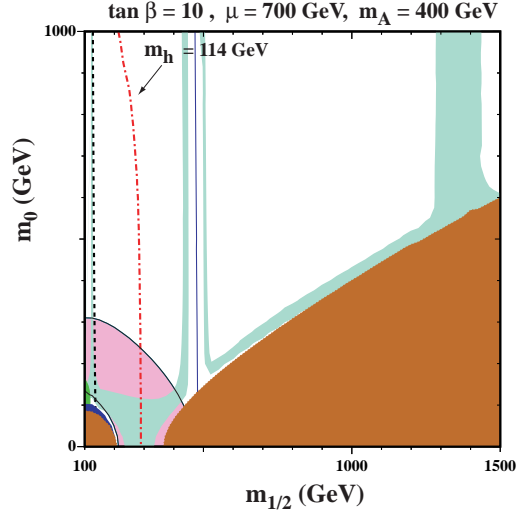}
}
\end{center}
\caption{\it The $(m_{1/2}, m_0)$ plane in the NUHM for $\tan \beta = 10$, 
$\mu = 700$~GeV and $m_A = 400$~GeV~\protect\cite{Ellis:2002iu}. The colours of the shadings
and contours are the same as in Fig.~\protect\ref{fig:UHM}.}
\label{fig:NUHM}       
\end{figure}

These cosmologically preferred regions move around
in the $(m_{1/2},m_0)$ plane if one abandons the universality assumptions of the
CMSSM. For example, if one allows the supersymmetry-breaking contributions
to the Higgs masses to be non-universal (NUHM), the rapid-annihilation WMAP 
`strip' can appear at different values of $\tan \beta$ and $m_{1/2}$, as seen in
Fig.~\ref{fig:NUHM}~\cite{Ellis:2002wv,Ellis:2002iu}. 
Rapid annihilation through the direct-channel $H, A$
poles suppresses the relic density between the two parallel vertical WMAP strips
at smaller values of $m_{1/2}$,
and the relic density is suppressed in the right-most strip because the neutralino
LSP has a significant higgsino component. A complete exploration of the parameter
space of the NUHM, which has two additional parameters compared to the CMSSM,
lies beyond the scope of this review.

%

\begin{figure}
\begin{center}
\resizebox{0.5\textwidth}{!}{%
  \includegraphics{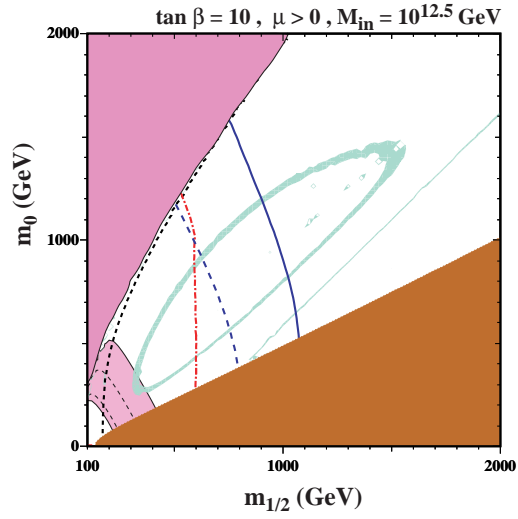}
}
\end{center}
\caption{The $(m_{1/2}, m_0)$ plane in a GUTless model with $\tan \beta = 10$,
$\mu > 0$  and $A_0 = 0$, assuming universality at $M_{in} = 
10^{12.5}$~GeV~\protect\cite{Ellis:2007ac}. The colours of the shadings
and contours are the same as in Fig.~\protect\ref{fig:UHM}.}
\label{fig:GUTless}       
\end{figure}

The appearance of the $(m_{1/2}, m_0)$ plane is also changed significantly if one
assumes that the universality of soft super-symmetry-breaking masses in the CMSSM
occurs not at the GUT scale, but at some lower renormalization 
scale~\cite{Ellis:2006vc,Ellis:2007ac,Ellis:2008ca}, as occurs in
some `mirage unification' models \cite{Choi:2005ge,Choi:2005uz,Falkowski:2005ck}. 
In this case, the sparticle masses are generally
closer together. As a consequence, the bulk, coannihilation,
rapid-annihilation and focus-point regions approach each other
and eventually merge as the mirage unification scale is reduced, as illustrated in
Fig.~\ref{fig:GUTless}, where they form an `atoll'. At smaller values of the mirage
unification scale, the atoll contracts and eventually disappears, and there is no
WMAP-compatible within the displayed portion of the $(m_{1/2}, m_0)$ plane.
In such `GUTless' models, $\Omega_{LSP} h^2$ falls below
the WMAP range (\ref{OCDM}) in larger regions of the $(m_{1/2}, m_0)$ plane
than in the conventional CMSSM with unification at the GUT scale.

%

\section{Summary}

As we have discussed above, there are many sound theoretical
and phenomenological reasons to favour supersymmetric
extensions of the Standard Model. In particular, supersymmetry predicts the existence of
cold dark matter in a very natural way, and there are several plausible
candidates for the lightest supersymmetric particle that would
be present as a relic from the Big Bang. The most
prominent candidate is the lightest neutralino, and we have described how
its relic density may be calculated, and the
regions of supersymmetric parameter space in which its density falls
within the range favoured by astrophysics and cosmology.
However, other candidates for the cold dark matter are
also possible, such as the gravitino. In that case, the next-to-lightest
supersymmetric particle would be metastable, and comparisons between
the observed light-element abundances and those predicted by
Big-Bang Nucleosynthesis calculations impose important constraints on
the parameter space. We have given examples of neutralino and gravitino
dark matter scenarios in the minimal supersymmetric extension of the
Standard Model, under various different theoretical assumptions. It will
be for collider and dark matter detection experiments to determine which,
if any, of these options has been adopted by Nature.

\subsection*{Acknowledgements}
The work of K.A.O.\ was partially supported by DOE grant DE-FG02-94ER-40823. 



\end{document}